\documentstyle[aps,psfig,multicol]{revtex}

\begin{document}
\title{Nonclassicality and information exchange in deterministic entanglement formation}
\author{M. C. de Oliveira$^1$\footnote{marcos@df.ufscar.br} and W. J. Munro$^2$}
\address{$^{1}$Departamento de F\'\i sica, CCET, Universidade Federal de S\~ao Carlos,\\
 Via Washington Luiz Km 235, S\~ao Carlos, 13565-905, SP, Brazil. \\
 $^2$Hewlett-Packard Laboratories, Filton Road,\\ Stoke Gifford,
 Bristol, BS34 8QZ, United Kingdom.}
\date{\today}

\maketitle

\begin{abstract}
We discuss the role of nonclassicality of quantum states as a necessary resource in
deterministic generation of
  multipartite entangled states. In particular for
three bilinearly coupled modes of the electromagnetic field, tuning
of the coupling constants between the parties allows the total system
to evolve into both Bell and GHZ states only when
 one of the parties is initially prepared in a nonclassical state.
A superposition resource is then converted into an
  entanglement resource.
\end{abstract}
\begin{multicols}{2}
\section{introduction}
Undoubtedly state entanglement of multipartite systems is one of
the most remarkable features of quantum mechanics, with both
fundamental and practical implications \cite{kwiat}. While
performing a central role in discussions of nonlocal correlations
\cite{epr}, it is also the basis for quantum cryptography
\cite{cryp}, teleportation \cite{tel} and dense coding \cite{dc}.
Nowadays, the most accessible and controllable source of
entanglement arises from the process of spontaneous parametric
down-conversion of photons in nonlinear crystals
\cite{kwiat,white}. Recently, experiments with trapped ions have
also demonstrated a high degree of control on entanglement
generation \cite{sackett}. Technological improvements on
experiments involving atoms and cavity fields in both optical
\cite{kimble} and microwave \cite{harochept} frequency regions may
lead to new scenarios for entanglement experiments involving
massive particles in the near future. Progress in this direction
was already achieved in the generation of atomic
Eistein-Podolsky-Rosen (EPR)
 pairs \cite{epr,hagley} and in the multiparticle entanglement
 engineering \cite{rauschenbeutel}.

 In quantum information theory, besides the challenge to characterize and quantify
 the amount of entanglement presented by multipartite systems (see e.g. \cite{munro}),
 there is a recent effort on understanding the entanglement
 capability of quantum operations \cite{cirac1}, and particularly the entanglement capability
 of (deterministic) unitary evolutions \cite{zalka,cirac2,smolin}. One of the main
 results
 of these investigations was the dependence of the degree of
 deterministic entanglement formation with the initial conditions, specially the available
 entanglement resource (the initial entanglement before unitary evolution).
Indeed, from a different point of view, it was recently
demonstrated that it is only possible to entangle two light beams
in a passive linear device (a beam splitter) if one of the input
beams is prepared in a nonclassical state
\cite{buzek,scheel,proof}. Remark that instead of the entanglement
resource (the initially available entanglement) the separate party
initial state resource (nonclassicality of one of the light beams
 state) played the role for entanglement formation.

For linear optics implementation of quantum information
processing, such as in \cite{milburnnature}, it would be valuable
to extend the above optical deterministic entanglement formation
to allow that  an available  nonclassical light source be directly
converted into an entanglement  resource following the protocol:
{\it Let an
   uncorrelated three partite system given by
  \begin{equation}
  |\Psi\rangle=|A\rangle\otimes|B\rangle\otimes|C\rangle,\end{equation}
  where the normalized state of the party $A$ is written as a superposition
  $|A\rangle=\sum_i\alpha_i|A_i\rangle$.
  We want an unitary evolution $U$ mapping the superposition content
  of $A$ to an entanglement content of $B+C$, while leaving $A$
  uncorrelated,\begin{equation}\sum_i\alpha_i|A_i\rangle\otimes|B\rangle\otimes|C\rangle
\longrightarrow |A^\prime\rangle\otimes
   \sum_i\alpha^\prime_i|B_i^\prime\rangle\otimes|C_i^\prime\rangle,\end{equation}
 where now $\alpha^\prime_i\propto\alpha_i$ are the Schmidt coefficients
 for the $B+C$ entangled state.}


 Throughout this paper we show the possibility to realize the above entanglement
  protocol
 in optical devices whenever a nonclassical light source is available.
 Particularly we show two optical implementations. One based on fields
 coupling in nonlinear crystals and another based on a
 beam-splitter arrangement.
 We begin in Sec. II remarking the ability to entangle light beams
through bilinear operations. In Sec. III we focus our attention on
a system with an infinite dimensional Hilbert space, a  three mode
linear bosonic system (linear equations of motion), corresponding
to a three partite continuous variable system.
 In Sec. IV we discuss the role of nonclassicality for the entanglement of the three modes.
 We assume the modes initially uncorrelated.
Classical states remain uncorrelated {as time evolves}. However
with a truly quantum resource or state (a superposition state) in
one of the modes we show how (near maximal) entangled  states are
formed (GHZ like).  It is then possible to convert the quantum
resource (nonclassicality) from one mode to the others entangling
them in a Bell like state. The mode with the initial quantum
resource is left unentangled following the protocol we addressed
above.
\section{Entangling power of bilinear operations}
The essential question posed here is to the entangling power of a
bipartite
 unitary operation $U$ given in terms of
 linear
Bogoliubov transformation maps- transformations whose generator
form a Lie algebra of the {\bf SU}($2$) group, whose generators
are of the form
\begin{equation}
\label{su2} \frac{1}{2}\left(ab^\dagger+a^\dagger b\right),\;\;
\frac{i}{2}\left(ab^\dagger-a^\dagger b\right),
\end{equation}
where $a$ and $b$ are annihilation operators for parties A and B,
respectively. Those transformations cannot map a classical state
into a nonclassical one. A bipartite bosonic system under the
above transformation,  such as two light beams on a beam-splitter
\cite{buzek,scheel}, can only be entangled if the
 state of, at least, one of the parties is nonclassical \cite{buzek,scheel,proof}.
As an example, let two uncorrelated light beams A and B prepared
in coherent states $|\alpha\rangle$ and $|\beta\rangle$ be
incident on the two arms of a beam splitter. Any coherent state
under this transformation evolves coherently and so the states of
the two output light beams A$_o$ and B$_o$ are also uncorrelated
$|\alpha^\prime\rangle$ and $|\beta^\prime\rangle$. However, if
one of the states of A or B is a legitimate quantum state, i.e., a
nonclassical state, then A$_o$ and B$_o$ will be left in an
entangled state \cite{buzek,scheel,proof}. Remark that this is not
due to the linearity of the transformation, since  the same is not
valid for linear Bogoliubov transformations of the form
\begin{equation}
\label{s11}
\frac{1}{2}\left(a^\dagger b^\dagger+a b\right),\;\; \frac{i}{2}\left(a^\dagger b^\dagger-a b\right),
\end{equation}
which together with (\ref{su2}) form the algebra of the sympletic
group {\bf Sp}(4,R)\cite{ekert}. Hamiltonians of the kind of
Eq.(\ref{s11}) are well known to generate scaling transforms -
stretching and contractions in such a way to conserve the volume
of the phase space, known in quantum optics as squeezing
\cite{braunstein}. It is straightforward to show that this
transformation map is able to entangle prepared in any initial
classical state. On the other hand the generators of the {\bf
SU}(2) group generates rotations in the system phase space.


While being impossible to entangle systems prepared in classical
states
 through linear Hamiltonians of the (\ref{su2}) form,
it is still possible to determine conditions for deterministic
generation of entanglement when one of the parties is prepared in
a special quantum state. This is quite useful as it allows one to
understand how a initial one-party quantum state is converted in a
$N$-parties quantum state.

Applicative examples of interacting systems following the
Hamiltonian above are easily found in quantum optics. It could
describe the interaction between light modes with distinct
polarization \cite{demartini}, the interaction of light fields and
the motion of trapped atoms as in the proposal of Parkins and
Kimble \cite{parkins}, or even the interaction between the ionic
motional degrees of freedom, let us say, in the {\it x, y and z}
directions, of a trapped ion \cite{steinbach}. In what follows we
give two possible implementations, describing the interaction of
three electromagnetic field modes in a nonlinear crystal
\cite{shih}, and three beams interacting through beam splitters
arrangement, such as depicted in Fig.1. We will initially consider
the general problem for three modes on a nonlinear crystal and
then at the end we specify the results for beams splitter
operations, where the time dependence is appropriately replaced by
transmission and reflection coefficients.

\section{Conversion of quantum resources in a nonlinear crystal}

We consider, as depicted in Fig. 1(a), a nonlinear crystal where
two classical pump fields of frequency $\nu$ and $\mu$ are
incident, each one suffering first a frequency conversion into an
idler and a probe field, and then the two idlers are combined in
an up-conversion process. Besides presenting entanglement in
frequency and polarization as discussed in Ref. \cite{shih}, this
system also presents entanglement in occupation number, as we
discuss below.  By appropriately considering time ordering of the
events and photon momentum conservation we can represent the
interaction of the two quantum modes, A, B, and C by an effective
Hamiltonian. The interaction part for the  frequency conversion
processes I and II is given by $ e^{-i\nu t}b_i^\dagger b+ e^{i\nu
t}b_i b^\dagger$ and $e^{-i\mu t}c_i^\dagger c+ e^{i\mu t}c_i
c^\dagger$, respectively, where $b$, $c$ are the annihilation
operators for the signal modes and $\nu$ and $\mu$ are the
classical pump fields frequencies. The process III of
up-conversion of the two idler modes $b_i$ and $c_i$ is given by $
b_i^\dagger c_i^\dagger a+ b_i c_i a^\dagger$. The last full
quantum Hamiltonian can be linearized assuming $b_i=
\beta_i+\delta b_i$ and $c_i= \gamma_i+\delta b_i$, where
$\beta_i$ and $\gamma_i$ are the idlers c-numbers variables. By
assuming only terms at most linear in $\delta b_i$ and $\delta
c_i$ and neglecting a constant pump $-\beta\gamma a+H.c$ effect,
we arrive at the effective Hamiltonian describing the interaction
between the three crystal output modes of EM field in the presence
of the two classical pumping as \cite{shih,yariv}
\begin{eqnarray}
H &=&H_{\rm free}+H_I
\end{eqnarray}
where
\begin{eqnarray}
H_{\rm free}&=&\hbar \omega _{a}a^{\dagger }a+\hbar \omega _{b}b^{\dagger }b
+\hbar \omega_{c}c^{\dagger }c\\
H_I&=&\hbar\lambda \left[e^{-i\left(\nu t-\phi\right)}a^\dagger b+
e^{i\left(\nu t-\phi\right)}a b^\dagger\right]\nonumber \\
&+&\hbar \kappa \left[e^{-i\left(\mu t-\theta\right)}a^\dagger c+
e^{i\left(\mu t-\theta\right)}a c^\dagger\right] .\label{hamil}
\end{eqnarray}
Here the coupling parameters $\lambda$ and $\kappa$ incorporate
the pump amplitudes, the coupling between the EM-field and the
crystal, integrated over the crystal volume, and the idler
c-number functions $\beta_i$ and $\gamma_i$. $\phi$ and $\theta$
incorporate phase of the above terms. This rotating wave
interaction is pictorially useful as it allows, as presented in
\cite{meu} for two modes,
 not only the transference of energy, but the fully
 transference of state between the modes.
\begin{figure}
 \centerline{$\;$\hskip 0
truecm\psfig{figure=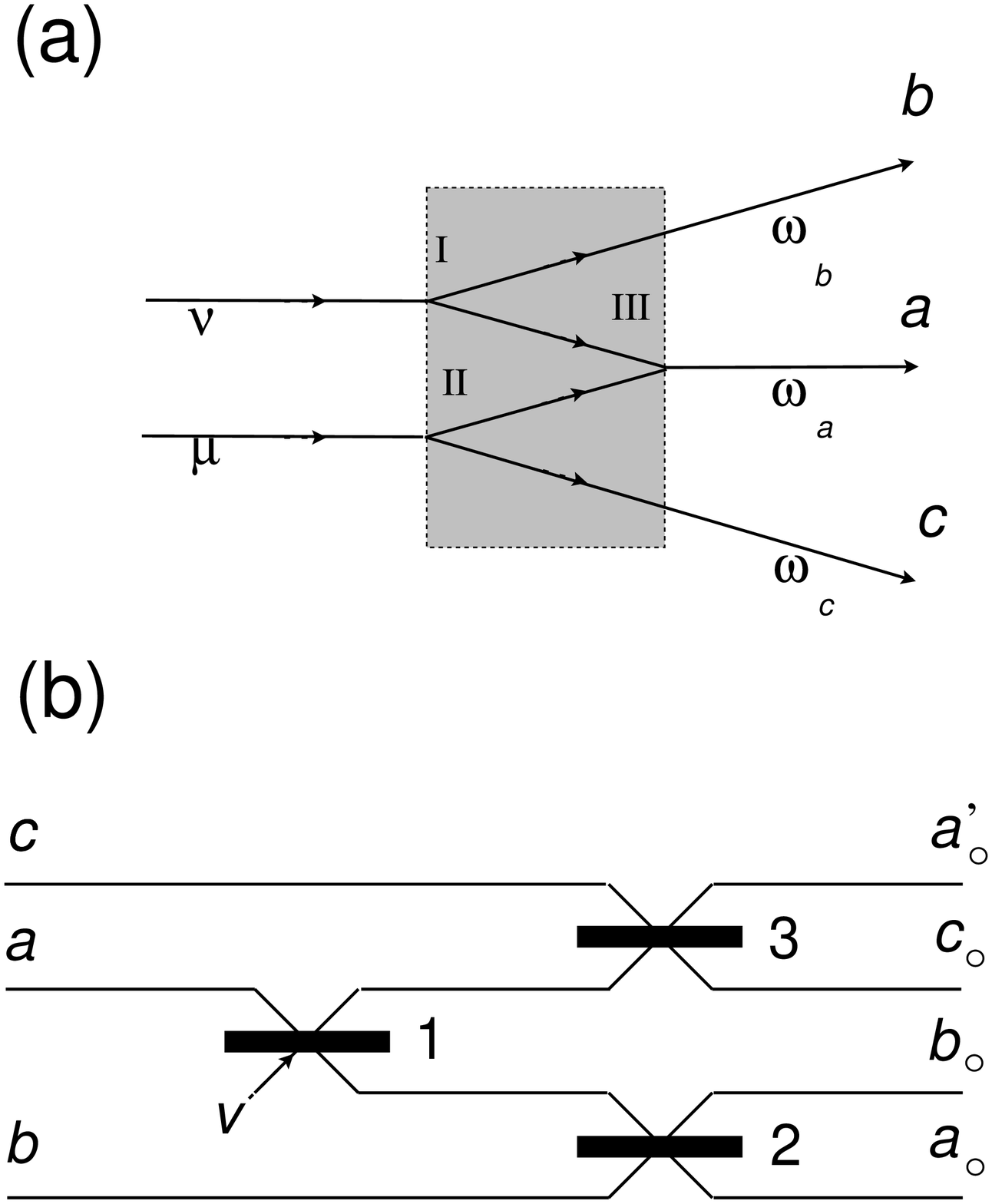,height=7cm}} 
\parbox{8cm}{\small Fig.1. Schematic setup for quantum resource conversion. (a)
 Nonlinear crystal output modes {\bf a}, {\bf b} and {\bf c}
are coupled by classical pump fields of frequency $\nu$ and $\mu$
in two frequency conversion (I and II) and a up-conversion (III)
processes. (b)In a beam-splitter arrangement the content of the
quantum state in the port {\bf a} is converted into entanglement
content of output ports {\bf b$_o$} and {\bf c$_o$}.} \label{fig1}
\end{figure}

The Heisenberg equations of motion for the field operators are given by
\begin{eqnarray} \label{equationsmotion}
\dot{a} &=&i \omega _{a} a -i\lambda
e^{i(\Omega t +\phi) }{b}-i\kappa e^{i(\Gamma t+\theta)} {c} \nonumber  \\
\dot{{b}} &=&i \omega _{b} b-i\lambda
e^{-i(\Omega t +\phi) }{a} \\
\dot{{c}} &=&i \omega _{c} c-i\kappa e^{-i(\Gamma t+\theta)} {a},
\nonumber
\end{eqnarray}
where $\Omega\equiv \omega_a-\omega_b-\nu$ and
$\Gamma\equiv\omega_a-\omega_c-\mu$. The equations of motion
(\ref{equationsmotion}) are linear and hence easily solvable.
Choosing $\Omega=\Gamma$ ($\omega_b+\nu=\omega_c+\mu$)
the solution for $a(t)$, $b(t)$ and $c(t)$ are
\begin{eqnarray}
a(t) &=& e^{-i\omega_a t} \left[u_{1}(t)a(0)+
v_{1}(t)b(0)+w_{1}(t)c(0)\right] \\
b(t) &=& e^{-i\omega_b t}
\left[u_{2}(t)a(0)+v_{2}(t)b(0)+w_{2}(t)c(0)\right] \\
c(t) &=& e^{-i\omega_c t} \left[u_{3}(t)a(0)+
v_{3}(t)b(0)+w_{3}(t)c(0)\right]
\label{bt}
\end{eqnarray}
where $a(0)$, $b(0)$ and $c(0)$ are the initial conditions for the
three modes. The time-dependent coefficients in the
above expression are given by
\begin{mathletters}\label{coeff}
\begin{eqnarray}
u_{1}(t)&=&e^{i\Omega t/2}f^*(t) \\
v_{1}(t)&=&-i\frac{\lambda e^{i\phi} e^{i\Omega t/2}}{A}\sin (At) \\
w_{1}(t)&=&-i\frac{\kappa e^{i\theta} e^{i\Omega t/2}}{A}\sin (At) \\
u_{2}(t)&=&-i\frac{\lambda e^{-i\phi} e^{-i\Omega t/2}}{A}\sin (At) \\
v_{2}(t)&=&1+\frac{\lambda^2}{\kappa^2+\lambda^2}\left[e^{-i\Omega
t/2}f(t)-1\right] \\
w_{2}(t)&=&\frac{\lambda\kappa e^{i(\theta-\phi)}}{\kappa^2+
\lambda^2}\left[e^{-i\Omega
t/2}f(t)-1\right] \\
u_{3}(t)&=&-i\frac{\kappa e^{-i\theta} e^{-i\Omega t/2}}{A}\sin (At) \\
v_{3}(t)&=&\frac{\lambda\kappa e^{-i(\theta-\phi)}}{\kappa^2+
\lambda^2}\left[e^{-i\Omega
t/2}f(t)-1\right] \\
w_{3}(t)&=&1+\frac{\kappa^2}{\kappa^2+\lambda^2}\left[e^{-i\Omega
t/2}f(t)-1\right]
\end{eqnarray} \end{mathletters}
where
\begin{mathletters}\begin{eqnarray}
f(t)&=&\left[\cos
(At)+\frac{i\Omega}{2A}\sin (At)\right], \\
A &=&\sqrt{\Omega^2+4(\kappa^2+\lambda^2)}/2.
\end{eqnarray}
\end{mathletters}

Using these expressions for the time-dependent Heisenberg
operators, it is possible via {\it characteristic functions} to
find expressions for the state vectors. The symmetric form of the
three modes characteristic function are given by
\begin{eqnarray}
\chi _{S}(\eta ,\zeta,\xi ,t)
&=&{\rm Tr}_{ABC}\left\{ \rho _{ABC}(0)\;\exp\left[\eta a^{\dagger
}(t)+\zeta b^{\dagger }(t)\right.\right. \nonumber \\
&+&\left.\left. \xi c^{\dagger }(t)-\eta ^{\ast }a(t)-\zeta ^{\ast
}b(t)-\xi ^{\ast }c(t)\right]\right\}
\label{chisim1}
\end{eqnarray}
where the RHS of the first (second) line stands for the Schr\"odinger
(Heisenberg) picture and $\rho_{ABC}$ is the density operator
for the three-mode system. As the  field operators
$a(t)$, $b(t)$ and  $c(t)$ (Heisenberg picture) depend linearly on
$a(0)$,  $b(0)$ and  $c(0)$ (Schr\"odinger picture)
we now define for convenience new time-dependent functions
\begin{eqnarray}
\bar{\eta} &\equiv& \eta u^\ast_1+\zeta u^\ast_2+\xi u^\ast_3,\\
\label{eta} \bar{\zeta} &\equiv& \eta v^\ast_1+\zeta v^\ast_2+\xi
v^\ast_3,\\  \label{zeta} \bar{\xi}&\equiv& \eta w^\ast_1+\zeta
w^\ast_2+\xi w^\ast_3. \label{xi}
\end{eqnarray}

\section{Initial condition for the entanglement of a three-partite bosonic system}
Before we move on with our discussion about the initial condition
role for deterministic entanglement formation we give a brief
review of three partite entanglement classification as given by
D\"ur {\it et al} \cite{dur},
 which we use extensively in what follows.  Let us write down four possible state configuration to which we will address later
\begin{eqnarray}
\rho(t)&=&\sum_i\left|a_i\right\rangle_A\left\langle a_i\right|\otimes
\left|b_i\right\rangle_B\left\langle b_i\right|\otimes\left|c_i\right\rangle_C\left\langle c_i\right| \label{d1}\\
\rho(t)&=&\sum_i\left|a_i\right\rangle_A\left\langle a_i\right|\otimes
\left|\psi_i\right\rangle_{BC}\left\langle \psi_i\right|\label{d2}\\
\rho(t)&=&\sum_i\left|b_i\right\rangle_B\left\langle b_i\right|\otimes
\left|\psi_i\right\rangle_{AC}\left\langle \psi_i\right|\label{d3}\\
\rho(t)&=&\sum_i\left|c_i\right\rangle_B\left\langle c_i\right|\otimes
\left|\psi_i\right\rangle_{AB}\left\langle \psi_i\right|\label{d4}
\end{eqnarray}
where,$\left|a_i\right\rangle$, $\left|b_i\right\rangle$ and $\left|c_i\right\rangle$ are (unnormalized) states of systems $A$, $B$, and $C$, respectively,
and $\left|\psi_i\right\rangle$ are states of two systems. One can classify a three-partite
state as:
The {\bf Class 1} ({\it Fully inseparable states}) is the one where the states cannot be written in any of the above forms- when the sate can be written in a state vector form.
 An obvious example is the GHZ $(\left|000\right\rangle+\left|111\right\rangle)/\sqrt{2}$, which is a maximally entangled state of three qubits.
 The {\bf Class 2} ({\it 1-qubit\footnote{
 Although the term {\it qubit} actually designates a superposition of a two-state system,
  we keep the original
 nomenclature, where the term is associated to the state of a party with
 an arbitrary Hilbert space dimension.}
 biseparable states}) constitutes the one where biseparable states with respect to one subsystem are states that are separable with respect to one subsystem,
 but non-separable with respect to the other two. That is, states that
 can be written in the form (\ref{d2}) or (\ref{d3}) or
(\ref{d4}), but not in both of them, i e, exclusively.
 The {\bf Class 3} ({\it 2-qubit biseparable states})
 relates to biseparable states with respect only to two
of the subsystems. Those are states that can be written in
 two and only two of the forms (\ref{d2}), (\ref{d3}) and (\ref{d4}).
 The {\bf Class 4} ({\it 3-qubit separable states}) are those
that can be written in all of the forms (\ref{d2}), (\ref{d3}) and (\ref{d4}),
but not as (\ref{d1}).
 Finally the {\bf Class 5} ({\it fully separable states}) is related
to states that can be written in the form (\ref{d1}).

Assuming that the initial joint density operator is
fully separable ({\bf class 5})\cite{dur},
\begin{equation}
\rho _{ABC}(0)=\rho _{A}(0)\otimes \rho _{B}(0)\otimes \rho _{C}(0) \, ,
\label{rho2}
\end{equation}
then the normal joint characteristic function (\ref{chisim1}) factorises as
\begin{eqnarray}
\chi _{N}(\eta ,\zeta,\xi ;t)&=&e^{\frac{1}{2}\left( \left|
\eta \right| ^{2}+\left| \zeta \right| ^{2}+
\left| \xi \right| ^{2}\right) }\chi_{S}(\eta ,\zeta,\xi ,t) \nonumber \\
&=&e^{\frac{1}{2}\left( \left|
\eta \right| ^{2}+\left| \zeta \right| ^{2}+\left| \xi \right| ^{2}\right) }
\chi _{S}(\bar{\eta},\bar{\zeta},\bar{\xi},0) \nonumber \\
&=&\chi _{N}^{A}(\bar{\eta};0)\;\chi _{N}^{B}(\bar{\zeta};0)\;\chi
_{N}^{C}(\bar{\xi};0) \,
\label{chiAchiB}
\end{eqnarray}
In the Fourier space the modes evolve in an apparent uncorrelated fashion.
The actual correlations are due to
the temporal functions $\bar{\eta}$, $\bar{\zeta}$ and $\bar{\xi}$.
To calculate the evolved joint quantum state we must assume specific initial
conditions.

\subsection{Coherent state}
The first and simplest situation to consider is the case where all the
modes are initially prepared in coherent states. Let us consider the
three modes to be given by the density operators
\begin{equation}
\rho _{A}(0)=\left|\alpha\right\rangle\left\langle\alpha\right|,
 \; \rho_{B}(0)=\left|\beta\right\rangle\left\langle\beta\right|,
 \; \rho _{C}(0)=\left|\gamma\right\rangle\left\langle\gamma\right|
 \end{equation}
where $\alpha$, $\beta$ and $\gamma$ are the amplitudes of the
respective coherent states. The normal ordered
characteristic function (\ref{chiAchiB}) can be
written as
\begin{eqnarray}
\chi _{N}(\eta ,\zeta,\xi ;t)
&=& e^{\eta x^*-\eta^*x+\zeta y^*-\zeta^* y+\xi z^*-\xi^* z},
\label{chiABC2}
\end{eqnarray}
where
 \begin{eqnarray} \label{xyz}
x&=& u_1\alpha+v_1\beta+w_1\gamma \nonumber \\
y&=&u_2\alpha+v_2\beta+w_2\gamma \\
z&=&u_3\alpha+v_3\beta+w_3\gamma \nonumber.
\end{eqnarray}
Comparing Eq. (\ref{chiABC2}) with the expression for the normal ordered
characteristic function in the Schr\"odinger picture, it is straightforward to write the density operator for the combined system as
\begin{equation}
\rho _{ABC}(t)=\left|\Psi_{ABC}(t)\right\rangle
\left\langle\Psi_{ABC}(t)\right|,
\end{equation}
where the joint state vector $\left|\Psi_{ABC}(t)\right\rangle$ is given by
\begin{equation}
\left|\Psi_{ABC}(t)\right\rangle=\left|x\right\rangle_{A}
\otimes\left|y\right\rangle_{B}\otimes\left|z\right\rangle_{C}.
\end{equation}
This  state vector simply shows that the three modes will never become
entangled. In fact they evolve as separate coherent states. We started
 with a fully separable quantum state and it evolved fully separable. The
amplitude of each of the coherent states does change,
but not its quantum signature.
While these states may develop classical correlations due to their
interaction, to observe quantum correlations such as entanglement
requires special conditions on the prepared states.
%

\subsection{Superposition states and quantum resources conversion}
This example raises several interesting questions. In what
follows, we still assume that the three modes are initially
uncorrelated but we consider the initial state of the mode A in a
nonclassical state - a superposition of coherent states. Does the
three modes remain uncorrelated as the total system evolves? How
is the information available in its state distributed over the
other modes? To begin this investigation let us describe the
initial states as
\begin{equation}
\rho _{A}(0)=\left|\psi_A\right\rangle\left\langle\psi_A\right|,
  \rho_{B}(0)=\left|\beta\right\rangle\left\langle\beta\right|,
  \rho _{C}(0)=\left|\gamma\right\rangle\left\langle\gamma\right|
\end{equation}
where mode $A$ is in the coherent superposition state
\begin{equation}
\left| \psi _{A}\right\rangle =\frac{1}{N^\alpha}\left(
\left| \alpha\right\rangle +e^{i\Phi}\left| -\alpha
\right\rangle \right),
\label{gato}
\end{equation}
with the normalisation factor $
N^\alpha = \sqrt{ 2+ 2\cos \Phi\; e^{-2|\alpha |^2}}$.
For $\Phi =0, \pi/2, \pi$ the superposition state given by (\ref{gato}) is
known as the {\em even cat state} (even coherent state) \cite{dodonov},
{\em Yurke-Stoler cat state} \cite{yurke}
and {\em odd cat state} (odd coherent state) \cite{dodonov}, respectively.

Now following the same steps as previously in the last example,
the final joint state of the three interacting modes is given by
\begin{eqnarray}\label{ghzstate}
\left|\Psi_{ABC}(t)\right\rangle&=&\frac{1}{N^\alpha}
\left(\left|X_1\right\rangle_A\otimes\left|Y_1\right\rangle_B
\otimes\left|Z_1\right\rangle_C\right.\nonumber\\
&+&\left.e^{i\Phi}\left|X_2\right\rangle_A\otimes
\left|Y_2\right\rangle_B\otimes\left|Z_2\right\rangle_C\right)
\end{eqnarray}
where
\begin{eqnarray} \label{xyz2}
\begin{array}{ccc}
X_1&=& u_1\alpha+v_1\beta+w_1\gamma,\; X_2= -u_1\alpha+v_1\beta+w_1\gamma \\
Y_1&=&u_2\alpha+v_2\beta+w_2\gamma,\; Y_2=-u_2\alpha+v_2\beta+w_2\gamma \\
Z_1&=&u_3\alpha+v_3\beta+w_3\gamma,\; Z_2=-u_3\alpha+v_3\beta+w_3\gamma\\
\end{array}\nonumber
\end{eqnarray}

In certain parameter regimes (namely $|\alpha|\gg 1$) we have
$\langle X_1| X_{2}\rangle\approx 0$,
$\langle Y_1| Y_{2}\rangle\approx 0$ and $\langle Z_1| Z_{2}\rangle\approx
0$. It can then be seen that (\ref{ghzstate}) is of
the form of an approximate GHZ state,
and so it is also very close to be maximally entangled
({\bf class 1})\cite{dur}.
Such characteristics come from the initial superposition
nature of the mode A state.

The evolution of these modes is also very interesting in other
parameter regimes and at specific times. For
simplicity we are going to restrict ourselves to the case in
which $\Omega=0$  ($\nu=\omega_a-\omega_b$ and $\mu=\omega_a-\omega_c$).
For $t=2n\pi/A$, where $n=$0, 1, 2..., it is easily
shown that the joint state recurs
to its initial uncorrelated form. Mode A
returns to a superposition state, while the modes B and C are coherent
states. No entanglement is seen for such times. However for times
given by
\begin{eqnarray}
t=\left(n-{1\over 2}\right) {\pi \over A}\;\;\;\;\;\;{\rm {n=1,2,3,..}}
\end{eqnarray}
the final joint state of the interacting modes is given by
\begin{eqnarray}
\label{fin}
\left|\Psi_{ABC}(t)\right\rangle&=&\frac{1}{N^\alpha}
\left|\bar X \right\rangle_A \nonumber\\
&\otimes&
\left(\left|Y_1\right\rangle_B\otimes\left|Z_1\right\rangle_C
+e^{i\Phi}\left|Y_2\right\rangle_B\otimes\left|Z_2\right\rangle_C\right)
\end{eqnarray}
where $\bar X= (X_{1}+X_{2})/2$. Here it is clear that mode A is
in a coherent state and is definitely not entangled with the modes
B+C. The initial superposition state content of the mode A is
completely converted in to an entanglement content (Schmidt
coefficient) of the couple BC. The resulting state is a {\it
1-qubit biseparable} ({\bf class 2}) entangled state \cite{dur}.
More specifically,
 when the modes B and C are prepared in vacuum state and
 with $\Phi=0$, the state of
 Eq. (\ref{fin}) is given by
\begin{eqnarray}
\label{fin2}
\left|\Psi_{ABC}(t)\right\rangle&=&\frac{1}{2N_+^\alpha}
\left|0\right\rangle_A
\otimes\left(N_+^{\lambda\alpha/A}N_+^{\kappa\alpha/A}\left|+
\right\rangle_B\otimes\left|+\right\rangle_C\right.\nonumber\\
&\;&\;\;\;\;\;\;\;\;\;+\left.N_-^{\lambda\alpha/A}N_-^{\kappa\alpha/A}
\left|-\right\rangle_B\otimes\left|-\right\rangle_C\right)
\end{eqnarray}
 where
\begin{eqnarray}
\left| \pm\right\rangle_B &=&\frac{1}{N_\pm^{\lambda\alpha/A}}
\left( \left| \lambda\alpha/A\right\rangle \pm
\left| -\lambda\alpha/A
\right\rangle \right) \\
\left| \pm\right\rangle_C &=&
\frac{1}{N_\pm^{\kappa\alpha/A}}\left( \left|
\kappa\alpha/A\right\rangle \pm\left| -\kappa\alpha/A
\right\rangle \right).
\end{eqnarray}
The couple B-C is in a truly nonseparable entanglement of
orthogonal states, independently of the magnitude of $\alpha$. In
fact when mode A is traced out from Eq. (\ref{fin2}), the partite
B+C system is left in a Bell state.

\section{ Quantum resource conversion with linear optical devices}

Linear optical devices offer an interesting scenario for
entanglement manipulation due the efficiency of linear processes
over nonlinear ones. Whenever a nonclassical light beam is
available, it can be manipulated together with other (classical)
beams through linear devices in order to achieve the required
operation. In what follows we discuss this situation for the
realization of the proposed protocol with three light beams and
beam-splitters. Consider the scheme of Fig. 1(b). The field  with
the quantum content, ($a$), is incident on the beam splitter 1.
The reflected ($a_r$) and transmitted ($a_t$) beams are then mixed
in beam splitters 2 and 3 with the fields $b$ and $c$,
respectively, to give the output beams $a_o$, $a_o^\prime$, $b_o$,
and $c_o$, through  the beam-splitter transformations
\begin{eqnarray}
a_t&=&T_1a+R_1v,\,\;\;\;\;a_r=T_1v+R_1 a\nonumber\\
a_o&=&T_2a_t+R_2b,\;\;\;\; b_o= T_2 b+R_2 a_t\nonumber\\
a_o^\prime&=&T_3 a_r+R_3c,\;\;\;\;c_o= T_3 c+R_3 a_r, \nonumber
\end{eqnarray}where $T_l=\cos\varphi_l$ is the complex
transmission coefficient of the beam-splitter $l$ and
$R_l=i\sin^2\varphi_l$ is its reflection coefficient, such that
$|T_l|^1+|R_l|^2=1$. $v$ is an annihilation operator representing
the input vacuum mode at beam-splitter 1. It is direct to show
that if the $a$ input field is prepared in a superposition of
coherent states as $|\psi\rangle=\sum c_j|\alpha_j\rangle$ while
the fields $b$ and $c$ are prepared in coherent states $\beta$ and
$\gamma$, respectively, the above beam-splitter transformation
will give\begin{eqnarray}
|\psi,0,\beta,\gamma\rangle_{in}&&\otimes|0,0,0,0\rangle_{out}\nonumber\\
&&\rightarrow
|0,0,0,0\rangle_{in}\otimes\sum_jc_j|\alpha_j^o,\alpha^{\prime
o}_j,\beta_j,\gamma_j\rangle_{out},\end{eqnarray} where we have
used the $\{a,v,b,c\}$ order for the input and the
$\{a_o,a_o^\prime,b_o,c_o\}$ order for the output, and
\begin{eqnarray}
\alpha_j^o&\equiv&\alpha_j T_1T_2+\beta R_2,\\
\alpha^{\prime o}_j&\equiv&\alpha_j R_1T_3+\gamma R_3,\\
\beta_j&\equiv&\alpha_j T_1R_2+\beta T_2,\\
\gamma_j&\equiv&\alpha_j R_1R_3+\gamma T_3.
\end{eqnarray}
If, for example we set $\varphi_2=\varphi_3=\pi/2$, such that
$R_2=R_3=i$ ($T_2=T_3=0$), the output will be \begin{equation}
|i\beta\rangle_{a_o}\otimes|i\gamma\rangle_{a_o^\prime}\otimes
\sum_jc_j|iT_1\alpha_j,iR_1\alpha_j+\rangle_{out},\end{equation}
and the quantum resource conversion will be complete.

\section{Conclusion}
 In summary, we have discussed in this article the possibility of conversion of
  quantum resources for the deterministic generation of multiparticle
  entangled systems.
  We have considered two realization examples of three bilinearly coupled modes of the
  electromagnetic field: (i) through interactions in a non-linear crystal, and
  (ii) through a
beam-splitter arrangement.
  Without the presence of
  a quantum resource (nonclassical state) for one of the three parties the resulting system
  remains completely unentangled (but may become classically
  correlated). If however one of the parties is initially prepared in a
  a nonclassical state, then this quantum
  resource can be transferred to the other parties. In fact a maximally
  entangled GHZ ({\bf class 1}) state for the three parties can be generated.
  As a central result the superposition resource has been converted into an
  entanglement resource. More interesting however is that this
  superposition resource for party A can be transferred completely to
  an entanglement resource for parties B and C only generating a {\bf class
  2} state. Parties B and C are left in a maximally entangled Bell state
  while party A is now an unentangled coherent state. A slightly
  different parameter regime leaves parties B and C strongly entangled
  to each other but weakly entangled to A while other parameter regimes
  leave the total system completely unentangled ({\bf class 5}). This results
  indicate the importance of the nature of the initial state when the
  modes are bilinearly coupled and how the nonclassical nature of one
  parties initial state can be shared or exchanged with other parties.
  A related inverse problem is the generation of superposition states. It is
straightforward to deduce from the above discussion that the
generation of a superposition (nonclassical) state in the mode A
is possible, in principle, whenever this mode interacts by the
Hamiltonian here discussed and if the two other modes B and C are
prepared in a entangled Bell state. It is interesting to notice
how these two intrinsic characteristic of quantum systems,
entanglement and superposition states are then related, and that
the conservation of quantum information is the strong bond that
relates them.
  Throughout this paper we have considered only superposition of coherent states as
  our quantum resource but any other genuine nonclassical
  states (for instance Fock states) generate similar effects.

\acknowledgments{
MCO is supported by FAPESP (Brazil) under projects $\#01/00530-2$ and $\#00/15084-5$
while WJM acknowledges funding in part by the European project EQUIP
(IST-1999-11053).}

%

 \end{multicols}
\end{document}